\documentclass[12pt,draft]{article}

\usepackage{amssymb}
\usepackage{latexsym}

\newtheorem{theorem}{Theorem}[section]

\newenvironment{proof}{\paragraph{\it Proof.}}{$\square$\vskip0.4cm}

\newcommand{\nc}{\newcommand}
\nc{\C}{{\mathbb C}}
\nc{\R}{{\mathbb R}}
\nc{\HH}{{\mathbb H}}
\nc{\Z}{{\mathbb Z}}
\nc{\dbar}{{\overline{\partial}}}
\nc{\dd}{{\rm d}}
\nc{\ii}{{\bf i}}
\nc{\jj}{{\bf j}}
\nc{\kk}{{\bf k}}
\nc{\qq}{{\bf q}}
\nc{\pp}{{\bf p}}
\nc{\ad}{\mathop{\rm ad}\nolimits}
\nc{\tr}{\mathop{\rm tr}\nolimits}
\nc{\su}{{\mathfrak s}{\mathfrak u}(2)}
\nc{\binomial}[2]{\mbox{\Large $#1 \choose #2$}}

\begin{document}

\thispagestyle{empty}

\title{Geometric interpretation of Schwarzschild instantons}

\author{ 
G\'abor Etesi
\\ {\it Yukawa Institute for Theoretical Physics,}
\\ {\it  Kyoto University,} 
\\{\it Kyoto, 606-8502 Japan,} 
\\ {\tt etesi@yukawa.kyoto-u.ac.jp}
\\
\\ Tam\'as Hausel 
\\ {\it Miller Institute for Basic Research in Science and}\\
{\it Department of Mathematics}
\\ {\it University of California at Berkeley} 
\\ {\it Berkeley CA 94720, USA}\\ 
{\tt hausel@math.berkeley.edu} }

\maketitle

\begin{abstract} In this note we address the problem of finding 
Abelian instantons of finite energy on the Euclidean Schwarzschild
manifold. This amounts to construct self-dual 
$L^2$ harmonic $2$-forms on the space. Gibbons found a non-topological $L^2$ 
harmonic form in the Taub-NUT metric, leading to Abelian instantons with 
continuous energy.
We imitate his construction in the case of the Euclidean Schwarzschild
 manifold and find a non-topological self-dual $L^2$ harmonic $2$-form on it.
 We show how this gives rise to Abelian instantons and identify them 
with $SU(2)$-instantons of Pontryagin number $2n^2$
found by Charap and Duff in 1977. Using results of Dodziuk and Hitchin we 
also calculate the full $L^2$ harmonic space for the
Euclidean Schwarzschild manifold. 
 \end{abstract}

\thispagestyle{empty}

\newpage

\section{Introduction}

An Abelian instanton is a self-dual solution to Euclidean Maxwell's
equations. In the case of the Taub-NUT metric on $\R^4$
such a non-trivial solution was found by Eguchi-Hanson \cite{eguchi-hanson}
in 1979. In mathematical terms a self-dual solution to Euclidean Maxwell's
equations with finite energy is a self-dual $L^2$ harmonic $2$-form with
integer cohomology class. In this context the 
Eguchi-Hanson solution was reinvented by Gibbons \cite{gibbons} in 1996. 
Motivated by Sen's S-duality conjecture he constructed a non-topological
\footnote{\label{footnote} In general we call a non-trivial 
$L^2$ harmonic form on a complete Riemannian 
manifold {\em non-topological} if either it is exact or 
not cohomologous to a compactly
supported differential form. Roughly speaking the existence of 
non-topological $L^2$ harmonic forms are not predictable by topological 
means. (Cf. \cite{segal-selby}.)} self-dual $L^2$ harmonic 
$2$-form in the Taub-NUT metric. A curious feature of this form is that, 
living on a space with no topology, it is
cohomologically trivial, producing a family of Abelian instantons with 
continuous energy. 

Gibbons' construction is geometric in nature; indeed the $L^2$ harmonic 
$2$-form is obtained 
as the exterior derivative of a $1$-form dual to a Killing field of some 
natural $U(1)$-action. In 1999 Hitchin \cite{hitchin} 
completed the proof of Sen's 
S-duality conjecture in the Taub-NUT case by showing that the {\em whole} 
$L^2$ harmonic space is spanned by the Eguchi-Hanson-Gibbons $2$-form. 

In this note we imitate this construction of Gibbons for the case of the 
Euclidean Schwarzschild metric. It is a Ricci-flat metric on $\R^2\times
S^2$ \cite{wald} and was constructed by Hawking \cite{hawking} in 1976 as 
the Wick rotation  of the Schwarzschild space-time. 

We show that the rotation on the $\R^2$ part 
induces a Killing field such that the exterior derivative of the dual 
$1$-form has finite energy. On a Ricci-flat manifold it follows from 
Killing's equations that the form obtained this way solves Maxwell's 
equations \cite{wald}. 
However, unlike the Taub-NUT case, this form is not self-dual (this fact
is related\footnote{Cf. Theorem 4 of \cite{hitchin}.} 
to the observation that the Euclidean Schwarzschild manifold is 
{\it not} hyperk\"ahler while the Taub-NUT manifold is). Self-dualizing
the form produces a self-dual $L^2$ harmonic $2$-form, which is {\em not}
trivial\footnote{Nevertheless it is still non-topological in the sense of  
 footnote \ref{footnote} above, since on $M$ every compactly supported 
$2$-form is exact.} 
 cohomologically. Thus in order to obtain Abelian instantons, we
have to quantize the form to have integer cohomology class. In this way we 
get   
Abelian instantons lying on $U(1)$-bundles of first Chern numbers $n$ and 
first Pontryagin numbers $2n^2$. 

On the other hand $SU(2)$-instantons on the Euclidean Schwarzschild
manifold were 
constructed by Charap and Duff \cite{charap-duff} in 1977. They considered 
$O(3)$-invariant instantons, where the action of $O(3)$ is induced from the 
symmetry group of $S^2$. In this way their ansatz was reduced to a system of 
three relatively simple partial differential equations. They were able to 
find
three kind of solutions of this system. The first was the trivial flat 
connection; the second the non-trivial ``metric connection'' of second
Chern number  $1$ obtained earlier in \cite{charap-duff1}; and the third 
was a family of  solutions which gave rise to instantons of second Chern
number $2n^2$. Apparently they refer to this last family as non-Abelian
dyons and give no geometrical interpretation.   

Representing $U(1)$ as a subgroup of $SU(2)$ we obtain $SU(2)$ instantons 
with second Chern numbers (i.e. instanton numbers) $2n^2$ from our integer
$L^2$ harmonic forms. The  main result of the present note is that this
family coincides with the third group of $SU(2)$-instantons found by Charap 
and Duff.  In spite of a few work dealing with or mentioning the
Charap--Duff instantons \cite{gibi}\cite{pope1}\cite{pope2} apparently its
Abelian character has not been recognized yet.

Using a recent result of Hitchin \cite{hitchin} 
we finish our paper by showing that there are no other Abelian instantons,
i.e. self-dual $L^2$ harmonic $2$-forms on the Euclidean Schwarzschild
manifold. Indeed with the help of a result of Dodziuk \cite{dodziuk} we 
are able to determine the {\em whole} $L^2$ harmonic space. 

\vskip.4cm 

\paragraph{\bf Acknowledgement.} The work in this paper was done when the 
second author visited the Yukawa Institute of Kyoto University in February 
2000. We are grateful for Prof. G.W. Gibbons for insightful discussions 
and Prof. H. Kodama and the Yukawa Institute for the invitation and
hospitality. 

\section{Construction of the Abelian instanton} Hawking invented the 
Euclidean Schwarzschild manifold to argue for the thermal nature of particle 
creation  at a Schwarzschild black hole. 

Mathematically the Euclidean Schwarzschild $4$-manifold $M$ 
is a complete solution to the Euclidean Einstein's equations with zero
cosmological constant, and has the non-trivial topology 
$M\cong \R^2\times S^2$. In other words it
is a Ricci flat manifold. It is not a gravitational 
instanton (such as e.g. the Taub-NUT metric or the Eguchi-Hanson metric) in
that its curvature tensor is not self-dual. Thus it is not hyperk\"ahler
either, which property will effect our considerations (cf. Theorem 4 of
\cite{hitchin}) in the form of the existence of non-self-dual 
$L^2$ harmonic forms on $M$.
 
According to (14.3.11) of \cite{wald},  
we have a particularly nice form of the metric $g$ on a dense open subset 
$(\R^2\setminus \{ O\})\times S^2\subset M\cong \R^2\times S^2$ of the
Euclidean Schwarzschild manifold. It is convenient to use polar
coordinates 
$(r,\tau)$ on $\R^2\setminus\{O\}$ in the range $r\in (2m,\infty)$ and 
$\tau\in [0,8\pi m)$, where $m>0$ is a fixed constant. The metric then 
takes the form 
\[\dd s^2=\left( 1-\frac{2m}{r}\right) \dd\tau^2+
\left( 1-\frac{2m}{r}\right)^{-1}\dd r^2+r^2\dd\Omega^2,\]
where $\dd\Omega^2$ stands for the line element of the unit round $S^2$. 
In  sphere coordinates $\Theta\in (0,\pi)$ and $\phi\in [0,2\pi)$ it is 
$$\dd\Omega^2=\dd\Theta^2+\sin^2\Theta\:\dd\phi^2$$ on the open coordinate
chart 
$(S^2\setminus(\{S\}\cup\{ N \} ))\subset S^2$.
Consequently the above metric takes the following form on the open, dense 
coordinate chart 
$U:=(\R^2\setminus\{O\})\times (S^2\setminus(\{S\}\cup\{N\}))\subset 
M\cong \R^2\times S^2$:
\begin{eqnarray} \dd s^2=\left( 1-\frac{2m}{r}\right)\dd\tau^2+
\left( 1-\frac{2m}{r}\right)^{-1}\dd r^2+r^2 (\dd\Theta^2+\sin^2\Theta
\dd\phi^2).
\label{metric}\end{eqnarray}
Despite the apparent singularity of the metric at the origin $ O\in\R^2$,
it can be extended analytically to the whole
$\R^2\times S^2$ as demonstrated on page 407 of \cite{wald}.

The $U(1)$-action defined by $\tau\mapsto \tau + 4m\lambda$ for 
$e^{i\lambda} \in U(1)$ leaves this metric invariant, and thus defines the 
Killing field 
\[ X:={1\over 4m}{\partial\over\partial \tau},\] 
which (together with 
the $U(1)$-action itself) clearly 
extends to a Killing field on the whole Euclidean Schwarzschild manifold,
which we will also denote by $X$. 

Now consider the differential $1$-form $\xi :=g(X,\:\cdot\:)$ 
dual to $X$. In our coordinate chart $U$ it takes the form 
$$\xi={1\over 4m}\left(1-{2m\over r}\right)\dd\tau .$$ 
General considerations about 
Killing's equations on a Ricci flat manifold yield that $\dd\xi$ is a 
harmonic $2$-form, which on a complete manifold is equivalent to saying that 
it is closed and coclosed. For a proof see page 442-443 of \cite{wald}. 
In our situation we can check it by hand that 
our form $$\dd\xi=-\frac{1}{2r^2}\dd\tau\wedge\dd r$$ 
is coclosed. For this we need to calculate $*\dd\xi$. Evoking the local
coordinate representation of the general Hodge operation
(e.g. page 5 of \cite{ble}), the Hodge-operation
$*\::\Omega^2(M)\rightarrow\Omega^2(M)$ on the Euclidean Schwarzschild 
manifold $(M, g)$ can be written as
\[*\dd\tau\wedge\dd r=r^2\sin\Theta\dd\Theta\wedge\dd\phi 
,\:\:\:\:\:*\dd\Theta\wedge\dd\phi
={1\over r^2\sin\Theta}\:\dd\tau\wedge\dd r ,\]
\[*\dd\tau\wedge\dd\Theta =-\left(1-{2m\over
r}\right)^{-1}\sin\Theta\dd r\wedge\dd\phi ,\:\:\:\:\:
*\dd r\wedge\dd\phi =-\left(1-{2m\over
r}\right){1\over \sin\Theta}\:\dd\tau\wedge\dd\Theta,\]
\[*\dd\tau\wedge\dd\phi=\left(1-{2m\over
r}\right)^{-1}{1\over\sin\Theta}\:\dd
r\wedge\dd\Theta ,\:\:\:\:\:*\dd r\wedge\dd\Theta=\left(1-
{2m\over r}\right)\sin\Theta\dd\tau\wedge\dd\phi .\]
The orientation is fixed such that $\varepsilon_{\tau r\Theta\phi}=1$.
From here we can see that $$*\dd\xi
=-{1\over 2}\sin\Theta\dd\Theta\wedge\dd\phi $$ is closed.
Thus $\dd\xi$ is indeed harmonic. Now we show that it is $L^2$
by calculating the Maxwell action of it:
using the parameterization of the Euclidean Schwarzschild manifold given
above we find  
\begin{eqnarray}
\Vert\dd\xi\Vert^2_{L^2(M)}=\Vert *\dd\xi\Vert^2_{L^2(M)}={1\over
8\pi^2}\int\limits_M \dd\xi\wedge *\dd\xi ={1\over
8\pi^2}\int\limits_0^{2\pi}\int\limits_0^\pi\int\limits_{2m}^\infty
\int\limits_0^{8\pi m}{\sin\Theta\over 4r^2}\dd
\tau\dd r\dd\Theta\dd\phi ={1\over 2}. 
\label{L2}
\end{eqnarray}
In this way we have produced a $2$-dimensional 
space of $L^2$ harmonic $2$-forms on $M$ spanned by $\dd\xi$ and
$*\dd\xi$, and a $1$-dimensional subspace of (anti)self-dual $L^2$
harmonic forms spanned by $\omega_\pm :=\dd\xi\pm *\dd\xi$. From now on,
without loss of generality we focus on self-dual forms only, i.e we
will use the notation $\omega:=\omega_+$. Hence the self-dual form
looks like
\begin{eqnarray}
\omega =-{1\over 2}\left({1\over r^2}\dd\tau\wedge\dd
r+\sin\Theta\dd\Theta\wedge\dd\phi\right)
\label{ehgorbulet}
\end{eqnarray}
on $U$. By (\ref{L2}), the Maxwell action or
$L^2$-norm of the self-dual $\omega$ is
given by
\begin{eqnarray}
\Vert\omega\Vert^2_{L^2(M)}={1\over
8\pi^2}\int\limits_M \omega\wedge\omega = {1\over
8\pi^2}\int\limits_M2\dd\xi\wedge *\dd\xi = 1. 
\label{energy}
\end{eqnarray}

The self-dual $2$-form $\omega$ is not trivial topologically; 
indeed its cohomology class can be easily identified with  the 
first Chern class of the $U(1)$-bundle $H$ whose
restriction $H\vert_{S^2}$ is nothing but the Hopf $U(1)$-bundle
(i.e. the positive generator of $H^2(S^2,\Z)$) through the 
isomorphism $H^2(\R^2\times S^2,\Z)\cong H^2(S^2,\Z)$ via the integral
\begin{eqnarray}-{1\over 2\pi}\int\limits_{S^2}\omega\vert_{S^2}=
-{1\over 2\pi}\int\limits_{S^2}*\dd\xi={1\over
4\pi}\int\limits_0^{2\pi}\int\limits_0^\pi\sin\Theta\dd\Theta\dd\phi = 1,
\label{egesz}\end{eqnarray} 
where we embedded $S^2$ into $M$ as $S^2\cong \{p\}\times S^2\subset 
M\cong \R^2\times S^2$, where for the sake  of simplicity $p\in \R^2$
differs from the origin. 

According to (\ref{egesz}) ${1\over 2\pi}\omega\in H^2(M,\Z)$ 
is an integer form, 
thus there is a connection $A_1$ on $H$, whose curvature
satisfies $F_{A_1}=\omega\kk$, where we used the identification 
${\mathfrak u}(1)\cong\kk\R$. Furthermore it is unique, since
$\pi_1(M)=1$, consequently any flat connection must be the trivial one. 
Similarly the $U(1)$-bundle $H^n$ admits a
unique connection $A_n$ such that $F_{A_n}=n\omega\kk$. 

Now we write down $A_n$ locally 
on two charts and explain how to glue them together: 
Let us denote by $H^\pm$ the northern and southern hemispheres of $S^2$
respectively, in other words $H^+$ is the set of points, where 
$\Theta\leq \pi/2$ and $H^-$ is the set, where $\Theta\geq \pi/2$. 
Consider the coordinate charts $U^\pm :=\R^2\times H^\pm$
of the space $M=\R^2\times S^2$. Clearly, $M=U^+\cup U^-$ and $U^+\cap
U^-\cong \R^2\times S^1$ is given by the points satisfying $\Theta =\pi
/2$.
 By integrating
(\ref{ehgorbulet}), in our coordinate
chart $U$ and an appropriate trivialization of $H^n$ the connection
$A_n$ takes the form ($c_1$, $c_2$ are arbitrary real constants): 
\begin{eqnarray*}A_n^\pm =\frac{n}{2}\left(\left( c_1-\frac{1}{r}\right)\dd\tau
+(c_2+\cos\Theta )\dd\phi\right)\kk.
\end{eqnarray*}
For this to extend to the North pole ($\Theta=0$) and respectively to the 
South pole ($\Theta=\pi$), we need to choose $c_2=-1$ on $U^+$ 
and respectively $c_2=1$ on $U^-$. Thus our connection $A_n$ takes the 
following shape on the charts $U^\pm$:
\begin{eqnarray}A_n^\pm =\frac{n}{2}\left(\left( c_1-\frac{1}{r}\right)\dd\tau
+(\mp 1+\cos\Theta )\dd\phi\right)\kk.
\label{local}
\end{eqnarray}
Note that these connection forms are regular along $U^+\cap U^-$ and are
related by the Abelian gauge transformation
\[A^+_n-A^-_n=-n\:\dd\phi\kk \]
given by $e^{-n\phi\kk}\in U(1)$ along $U^+\cap U^-$. We
recognize the above connections as the $L^2$ harmonic generalizations
for the Euclidean Schwarzschild case of the connections appearing in the
well-known bundle-theoretic description of the Dirac magnetic monopole,
see e.g. page 231-232 of \cite{egu-gyl-han}. The extra term
$(c-1/r)\dd\tau$ can be interpreted as a scalar potential and will cause
that our solutions carry electric charge.

Consider now the associated
$U(2)$-bundle $P_{U(2)}\cong H^n\oplus H^{-n}$, 
via the diagonal embedding of 
$U(1)\times U(1)\subset U(2)$, and the associated connection 
$B_n=A_n\oplus A_{-n}$ with curvature
form $F_{A_n}\oplus F_{A_{-n}}$ on it.  Since $H^4(M,\Z)\cong 0$
the principal $U(2)$-bundle $P_{U(2)}$ of $H^n\oplus H^{-n}$ is
trivial. Moreover its determinant $U(1)$-bundle is trivial and thus 
$P_{U(2)}$ reduces to the trivial $SU(2)$-bundle which we
denote by $P = M\times SU(2)$.  
Furthermore the $U(2)$-connection 
$B_n$ induces a 
trivial connection on the  
determinantal $U(1)$-bundle so it reduces 
to an $SU(2)$-connection on $P$. In our
coordinate charts $U^\pm$ the connection 
$B_n$ is induced by the embedding $\kk\R\cong{\mathfrak u}(1)\subset\su
\cong{\rm Im}\HH$. In other words  self-dual $L^2$ harmonic $2$-forms may 
be regarded as the curvature
$2$-forms of (reducible) self-dual Yang-Mills $SU(2)$-connections given 
locally by the formula (\ref{local}). 

Using (\ref{energy}) we find that the second Chern numbers of these
self-dual Yang-Mills $SU(2)$-connections $B_n=A_n\oplus A_{-n}$ on the 
associated $SU(2)$-bundles $H^n\oplus H^{-n}$ satisfy: 
\[-{1\over 8\pi^2}\int\limits_M\tr\left( F_{A_n}\oplus
F_{A_{-n}}\wedge F_{A_n}\oplus F_{A_{-n}}\right)={1\over
8\pi^2} \int\limits_M 2F_{A_n}\wedge F_{A_n}=2n^2\]
since we have $-\tr (AB)=2{\rm Re}(x\overline{y})$ for the Killing-form on
the Lie algebra $\su\cong{\rm Im}\HH$. 

Note that if we calculate the first Pontryagin number of the connection
$A_n$ on the real plane bundle $H^n$ 
(here we made the identification $U(1)\cong SO(2)$) 
we also find
\[{1\over 4\pi^2}\int\limits_M F_{A_n}\wedge F_{A_n} = 2n^2.\]
In the following section we prove that the reducible $SU(2)$-instantons just
derived coincide with the third group of instantons  found by 
Charap and Duff \cite{charap-duff}.

\section{Identification with instantons of Charap and Duff}

Now we will follow \cite{charap-duff}. In that paper solutions of type  
(II) of the self-duality equations on $P$ 
are referred to as "non-Abelian dyons"
of Pontryagin numbers $2n^2$. Let us denote them $\widetilde{A}_n$. 
In this section we show that they are in fact  
reducible, i.e.  
Abelian connections and identify them with the connections $B_n$ 
defined above. To round things off, we finish this section by giving 
the explicit local gauge transformations which identify our Abelian 
connections (\ref{local}) with Charap--Duff's (\ref{connection}).

Let $n$ be an integer 
and focus our attention to solution (II), more precisely the self-dual
one, which means that we choose all the functions of positive sign.
Putting solution (II) into the spherical symmetric ansatz (5) of
\cite{charap-duff} and adjusting notations of \cite{charap-duff} 
to ours via the identification $\su\cong{\rm Im}\HH$ given by 
$\{\sigma^1/2,\sigma^2/2,\sigma^3/2\}\mapsto\{\ii /2 ,\jj /2 ,\kk /2\}$,
the coordinate transformation 
\begin{eqnarray}
(\tau , x^1, x^2, x^3)\longmapsto (n\tau, r\sin\Theta\cos (n \phi)
,r\sin\Theta\sin(n \phi) ,r\cos\Theta )
\label{transformation}
\end{eqnarray} and the notation 
$$\qq_n:=\sin\Theta\cos(n\phi)\ii
+\sin\Theta\sin(n\phi)\jj +\cos\Theta\kk,$$
we get the new form for the self-dual connection 
\begin{eqnarray}
\widetilde{A}_n={n\over 2}\left(\left(
c-{1\over r}\right)\dd\tau +\cos\Theta\:\dd\phi\right)\qq_n 
-{n\over 2}\dd\phi\kk 
+{1\over 2}\dd\Theta (\sin (n\phi )\ii -\cos (n\phi )\jj ).
\label{connection}\end{eqnarray}
A long but 
straightforward calculation shows that the curvature takes the form
\begin{eqnarray}
{F}_{\widetilde{A}_n}=n \omega\qq_n. 
\label{gorbulete}
\end{eqnarray}

Consider now the $U(1)$-sub-bundle $H_n$ of $P$ whose
smooth sections are given by 
$s=\exp(f \qq_n ),$ where $\exp:\su\rightarrow SU(2)$ is
the exponential map and $f$ is any smooth function on $M$. 
We show that the covariant derivative
$\nabla_{\widetilde{A}_n}:\Omega^0(\ad(P))\rightarrow\Omega^1(\ad(P))$
on the associated bundle $\ad (P)$
leaves the real line bundle $\ad(H_n)\subset \ad(P)$ invariant. We
thus calculate in our coordinate chart $U$:
\begin{eqnarray*} 
\nabla_{\widetilde{A}_n} s =
 \nabla_{\widetilde{A}_n} (f \qq_n )& = & \dd ( f
\qq_n)+ 
\left[\widetilde{A}_n,f\qq_n ,\right]
\end{eqnarray*} 
where by abuse of notation $\widetilde{A}_n$ denotes the connection
matrix of  
$\widetilde{A}_n$ in the gauge (\ref{connection}).
The first term equals:
\begin{eqnarray*}
\dd (f\qq_n) & = & \dd f\qq_n   +  f \dd
(\sin\Theta\cos(n \phi)\ii
+\sin\Theta\sin( n \phi)\jj +\cos\Theta\kk)\\ & = & 
\dd f\qq_n  +
f\dd\Theta\left(\cos\Theta\cos(n\phi)\ii+\cos\Theta\sin(n
\phi)\jj-\sin\Theta\kk\right)+\\
&& + \ \ 
fn\dd\phi\left(-\sin\Theta\sin(n\phi)\ii+\sin\Theta\cos(n\phi)\jj\right),  
\end{eqnarray*}
and the second gives,
\[\left[\widetilde{A}_n\:,\:f\qq_n \right] =\]
\[=\left[ {n\over 2} \left(
\left( c-{1\over r}\right)\dd\tau +\cos\Theta\:\dd\phi\right)\qq_n
- {n\over 2}\dd\phi\kk +{1\over 2}\dd\Theta (\sin(n \phi)\ii
-\cos(n\phi)\jj )\:, \:f\qq_n \right] =\] 
\[=\left[-{n\over 2}\dd\phi\kk +{1\over 2}\dd\Theta \left(\sin(n \phi)\ii
-\cos(n \phi)\jj\right)\:,\:f\qq_n \right]=\] 
\[ =fn\dd\phi\left(\sin\Theta\sin(n\phi)\ii - 
\sin\Theta\cos(n \phi)\jj\right)+ \]
\[+f\dd\Theta\left(-\cos\Theta\cos (n\phi)\ii-\cos\Theta\sin (n\phi)\jj + 
\sin\Theta\kk\right).\] 
Adding the two above expressions up we see that 
$$ \nabla_{\widetilde{A}_n} (f \qq_n)= \dd f \qq_n, $$
showing that $\widetilde{A}_n$ reduces to a $U(1)$-connection on 
$H_n\subset P$. Now (\ref{gorbulete}) shows that 
this $U(1)$-connection on $H_n$ has the
same curvature than $A_n$ therefore they should coincide, in particular 
$H_n\cong H^n$. Thus we proved that the Charap-Duff's connection 
(\ref{connection}) is equivalent to our connection (\ref{local}).

We finish this section by writing down the explicit gauge 
transformations on $U^\pm$ which transform our connection (\ref{local}) to 
Charap-Duff's (\ref{connection}). From (\ref{gorbulete}) we can guess
that the gauge transformations we are looking for
should rotate the vector $\qq_n$ 
into the unit vector $\kk$. This transformation
cannot be carried out continously over the whole $S^2$ by using only one
transformation but there is no obstruction if we use two gauge 
transformations on the charts $U^\pm$  which are related along $U^+\cap
U^-$ by an Abelian gauge transformation. Consider the gauge transformations
$g_n^\pm :U^\pm\rightarrow SU(2)\cong S^3\subset\HH$ given by\footnote{By
abuse of notation we will regard the unit quaternions $\ii ,\jj ,\kk$
either elements of the Lie algebra $\su\cong {\rm Im}\HH$ or of the group
$SU(2)\cong S^3\subset\HH$ depending on the context.}
\[ g^\pm_n(\tau, r,\Theta,\phi ):=\exp\left(\pm\kk{n\phi\over
2}\right)\exp\left(-\jj{\Theta\over 2}\right)\exp\left(-\kk{n\phi\over
2}\right) .\]
In this form we only  see that $g_n^\pm$ are smooth gauge transformations
on $U$. In order to be well defined as smooth 
maps $g_n^\pm :U^\pm\rightarrow SU(2)$ we have to show that they extend 
analytically over the appropriate poles. We show this for $g^+$ here, the
case of $g^-$ being similar. It is easily checked 
that the following gauge transformation in  Descartes coordinates gives 
rise\footnote{In other words the gauge transformation 
(\ref{descartes}) pulls back to $g^+$
by the map given by (\ref{transformation}).}
to $g^+$ after the coordinate 
transformation (\ref{transformation}):
\begin{eqnarray}
\left( {x_3\over 2r}+{1\over 2}\right)^{-1/2}\left( {x_3\over 2r}+{1\over 2}-
\ii {x_2\over 2r}-{\jj }{x_3\over 2r}\right).\label{descartes}\end{eqnarray}
In this form we see that the map $g^+:U\rightarrow SU(2)$ extends
analytically to $U^+\setminus U$, that is to 
points of $M$, where $(\Theta=0)$ or equivalently $ x_3/r =1$. 

Let us prove that the above gauge transformations do indeed transform
(\ref{connection}) into (\ref{local})!
First we show that it rotates $\qq_n$ into $\kk$: 
Writing  $\qq_n=\sin\Theta\cos (n\phi )\ii +\sin\Theta\sin (n\phi )\jj
+\cos\Theta\kk =\exp(\kk n\phi )\sin\Theta\ii +\cos\Theta\kk$, we can
proceed as follows:
\[g^\pm_n(\exp (\kk n\phi )\sin\Theta\ii +\cos\Theta\kk )(g_n^\pm)^{-1} 
=\]
\[=\exp\left(\pm\kk{n\phi\over 2}\right)\exp\left(-\jj{\Theta\over
2}\right)(\sin\Theta\ii +\cos\Theta\kk )\exp\left(\jj{\Theta\over
2}\right)\exp\left(\mp\kk{n\phi\over 2}\right) .\]
Since $\sin\Theta\ii +\cos\Theta\kk =\exp (\jj\Theta )\kk$, we can go
further by writing
\[\exp\left(\pm\kk{n\phi\over 2}\right)\kk\exp\left(\mp\kk{n\phi\over
2}\right)=\kk\] 
proving that the above gauge transformations $g^\pm_n$
send $\qq_n$ into $\kk$.

Finally we calculate that at one hand 
\[g^\pm_n\dd (g^\pm_n)^{-1} = \mp{n\over 2}\dd\phi\kk +{n\over 2}\dd\phi
\exp\left(\pm\kk{n\phi\over 2}\right)\exp (-\jj\Theta
)\exp\left(\mp\kk{n\phi\over 2}\right)\kk +{1\over 2}\dd\Theta\exp
(\pm\kk n\phi )\jj ,\]
on the other hand
\[g_n^\pm\left( -{n\over 2}\dd\phi\kk
+{1\over 2}\dd\Theta (\sin (n\phi )\ii -\cos (n\phi )\jj
)\right)(g^\pm_n)^{-1} =\]
\[= -{n\over 2}\dd\phi\exp\left(\pm\kk{n\phi\over
2}\right)\exp (-\jj\Theta )\exp\left(\mp\kk{n\phi\over 2}\right)\kk
-{1\over 2}\dd\Theta\exp (\pm\kk n\phi )\jj .\] 
But these terms cancel each other except $\mp {n\over
2}\dd\phi\kk$ demonstrating the desired result
\[g^\pm_n\widetilde{A}_n(g^\pm_n)^{-1} +g^\pm_n\dd (g^\pm_n)^{-1} =
A^\pm_n\]
where $A^\pm_n$ are given by (\ref{local}). Note that the two gauge
transformations are related along $U^+\cap U^-$ by the Abelian
gauge transformation
\[\exp (\kk n\phi ) g^-_n=g^+_n \]
yielding again $A^-_n -\kk n\dd\phi = A^+_n$.

Thus we gave two proofs that the Charap--Duff instantons coincide with 
ours proving that these solutions are nothing but Abelian dyons carrying
magnetic charge $n$ and electric charge $n$. Indeed, the electric
charge is given by the integration of the electric field over an embedded
two-sphere. By self-duality
\[-{1\over 2\pi}\int\limits_{S^2}*\omega\vert_{S^2}=1,\]
hence it is clear that the general solution has electric charge $n$, too. 
In summary we see that the basic
characteristic numbers of these solutions are their magnetic charge $n$
represented by the first Chern class of the $U(1)$-bundle 
$H^n$ instead of the
first Pontryagin number $2n^2$. 

\section{$L^2$-cohomology}
In this final section we show that we have found all the Abelian
instantons on the Euclidean Schwarzschild manifold. 

\begin{theorem} Let $\eta$ be an $L^2$ harmonic form on $M$. Then it is a 
linear combination of $\dd \xi $ and $*\dd \xi$. Consequently a self-dual
$L^2$ harmonic $2$-form on $M$ is some constant multiple of 
$\omega =\dd\xi + *\dd\xi$. 
\end{theorem}

\begin{proof} First of all the volume of $(M,g)$ is infinite. It can be
seen by calculating:
\[\int\limits_M*{\bf 1}
=\int\limits_0^{2\pi}\int\limits_0^\pi\int\limits_{2m}^\infty\int\limits_0^
{8\pi m}r^2\sin\Theta\dd\tau\dd r\dd\Theta\dd\phi =\infty,\] 
where we have used again the parameterization of the Euclidean
Schwarzschild manifold given in the previous section. 
This implies that there are no $L^2$ harmonic $0$- or equivalently
$4$-forms. Now, as $M$ is Ricci-flat and complete, 
Corollary 1 of Dodziuk \cite{dodziuk} implies that there
are no $1$- and equivalently $3$-forms on $M$. 

It remains to show that any $L^2$ harmonic $2$-form is a linear combination 
of $\dd\xi$ and $*\dd\xi$. For this we use a recent result of Hitchin,
namely Theorem 3 of
\cite{hitchin} which we cite in full:

\begin{theorem}[Hitchin] Let $M$ be a complete oriented Riemannian manifold
and let $G$ be a connected Lie group of isometries such that the Killing 
vector fields $X$ it defines satisfy $$\vert X\vert\leq c^\prime
\rho(x_0,x)+ c^{\prime\prime}.$$ Then each $L^2$ cohomology class is fixed
by $G$.
\end{theorem}
(Here $\rho$ is the distance function of the Riemannian manifold.)
We would like to apply this result to $M$ with $G\cong SO(3)$ acting on $M$ 
by isometries of $S^2$. A glance at the metric (\ref{metric}) assures us
that the Killing fields of this action have indeed linear growth. 
Thus it is sufficient to  
find all $SO(3)$-invariant harmonic $2$-forms on $M$. Let $\eta$ be such
a form. In our coordinate chart $U$ it  must have the shape: 
$$\eta=f(\tau, r)\dd\tau\wedge\dd r + \alpha_\tau (\tau ,r)\wedge\dd\tau +
\alpha_r(\tau ,r)\wedge\dd r +\beta (\tau ,r),$$ 
where $f(r,\tau)$ is an $SO(3)$-invariant function on $S^2$, moreover 
$\alpha_\tau (\tau ,r)$ and $\alpha_r(\tau, r)$ are $SO(3)$ invariant
$1$-forms on $S^2$, and
finally $\beta (\tau ,r)$ is an $SO(3)$-invariant $2$-form on $S^2$.
However there are very few $SO(3)$-invariant forms on $S^2$.
Namely only the constant functions and constant times the volume form of the 
round $S^2$ are $SO(3)$-invariant. It follows because $SO(3)$ acts 
transitively showing that only the constant functions and equivalently 
constant multiples of the volume form are the $SO(3)$-invariant $0$- and 
$2$-forms respectively. Moreover there are no non-trivial
$SO(3)$-invariant $1$-forms on $S^2$, which could be seen by looking at
the dual vector field and seeing that the action of the $U(1)$
stabilizator of any point on the tangent space at that point has only 
the origin as its fixed point. 

It follows that our $SO(3)$-invariant $2$-form must have the form:
$$\eta=f(\tau ,r)\dd\tau\wedge \dd r + h(\tau
,r)\sin\Theta\dd\Theta\wedge\dd\phi$$ where 
$-\sin\Theta\dd\Theta\wedge \dd\phi$ is the volume form  of the
unit $S^2$ and $f(\tau ,r)$ and $h(\tau ,r)$ stand for a 
function on $M$ depending  only on $\tau$ and $r$. Its Hodge-dual is
given by $$*\eta = h(\tau ,r)\frac{1}{r^2}\dd\tau\wedge\dd r
+r^2f(\tau , r)\sin\Theta\dd\Theta\wedge\dd\phi .$$
In order that both $\eta$ and $*\eta$ be closed we need that neither 
$h(r, \tau )$ nor $r^2f(r, \tau )$ depend on $\tau$ or $r$ which means
that $\eta$ must have the form: 
$$\frac{c_1}{r^2}\dd\tau\wedge\dd r+ c_2\sin\Theta\dd\Theta\wedge\dd\phi
,$$ exactly as claimed. The result follows. 
\end{proof}

\section{Concluding Remarks}

Previously we have proved that the self-dual solutions to the $SU(2)$
Yang--Mills equations over the Euclidean Schwarzschild manifold found by
Charap and Duff correspond to Abelian dyons rather than non-Abelian ones.
From the mathematical point of view we have seen that the curvatures of
these solutions represent elements of the non-trivial second reduced $L^2$
cohomology group of the Euclidean Schwarzschild manifold. This
identification enabled us to find all the Abelian instantons over this
manifold.

The physical interpretation of these solutions is more subtle, however. In
light of our results these solutions seem to describe a static  
electromagnetic dyon configuration surrounding the Schwarzschild black
hole. Accepting this, we can interpret their Pontryagin numbers $2n^2$ as
their three dimensional energy rather then their Euclidean action. Indeed,
it is straightforward that the Euclidean Schwarzschild metric tends to the
{\it three} dimensional flat metric of $\R\times S^2$ and can be
extended as the flat metric to the whole $\R^3$ as $m\rightarrow 0$
(i.e. as the Hawking temperature of the black hole tends to infinity)
while neither solutions (\ref{local}) nor their Euclidean action 
depends on $m$. Henceforth in the limit $m\rightarrow 0$ we recover the
static dyon of charge $(n,n)$ on flat space and such a configuration has
energy $2n^2$ as it is well known. 

The general (non self-dual) dyons of charge $(k, n)$ correspond to the
general elements of the integer lattice
$\Z\oplus\Z\subset\R\oplus\R\cong\overline{H}^2_{L^2}(M,g)$ in the reduced
$L^2$-cohomology group of $(M, g)$.

\end{document}